\documentstyle[openbib,12pt,fleqn]{article}
\newcommand{\bea}{\begin{eqnarray}}
\newcommand{\beq}{\begin{equation}}
\newcommand{\eea}{\end{eqnarray}}
\newcommand{\eeq}{\end{equation}}

\renewcommand{\baselinestretch}{2}

\begin{document}
\bibliographystyle{unsrt}

\setcounter{footnote}{0}

%
%
%
%
\begin{center}
\phantom{.}
{\Large \bf Effect of noise for two interacting particles in
a random potential \\}
{\small \sl   F.~BORGONOVI~$^{[a]}$\\}
{\small \it Dipartimento di Matematica, Universit\`a Cattolica,
via Trieste 17, 25121 Brescia, ITALY}

{\small  \sl D.L.~SHEPELYANSKY~$^{[b]}$ 
\\}
{\small \it Laboratoire de Physique Quantique,
  Universit\'e Paul Sabatier, \\
  118, route de Narbonne,31062 Toulouse Cedex, FRANCE}\\

\vspace{0.5truecm}

\vskip .3 truecm

\vspace{0.5truecm}
\end{center}
\small
{\bf Abstract:\/}
We investigated the effect of noise on propagation
of two interacting particles pairs in a quasi one--dimensional
random potential. It is shown that pair diffusion 
is strongly enhanced by short range interaction
comparing with the non--interacting case.
\vskip .6truecm

{PACS numbers {71.55.Jv, 72.10Bg, 05.45.+b }}
\newpage

Recent investigations showed that two interacting particles 
(TIP) in a random  potential can propagate coherently on a distance
$l_c$ which is much bigger than the one--particle localization length
$l_1$ without interaction \cite{TIP,Imry,Pichard,Borg1,Oppen,Frahm}. 
According to \cite{TIP,Imry} the TIP localization length is given by:
\bea
l_c \sim l_1^2 M (U/V)^2/32 
\label{tipe}
\eea
where $M$ is the number of transverse channels in a quasi one--dimensional 
wire, $U$ is the strength of on-site interaction 
assumed to be comparable to or less than the bandwidth
and $V$ is the 
hopping matrix element between nearby sites.
Here it is also assumed that the wave vector $k_F$ corresponding
to TIP energy is $k_F \sim 1/a = 1$, $a$ being the lattice constant.
While the exact verification of (\ref{tipe}) is still under investigation
the existence of the enhancement of
two particle localization length have been clearly demonstrated  
in numerical simulations
\cite{Pichard,Borg1,Oppen,Frahm}.
These simulations have been done for different models. 
Main results are for 1--d  Anderson model with TIP 
\cite{Pichard,Oppen}
and  
for the model of two interacting kicked rotators \cite{Borg1}.
This last  model is very convenient for the investigation 
of wave packets spreading in time due to the existing effective
numerical methods.
For this reason, in our numerical studies, we used the last model.

The problem we want to address in this letter is the influence of 
noise on TIP localization.
For one particle the effect of noise has been analyzed during
the last years and the physics of this phenomenon is well understood
\cite{Ott,Doron,Fishman,Borg2}.
Generally, the main effect of noise is the destruction of interference
after a coherence time $t_c$, after which the particle makes a jump 
on a distance $l_1$ that leads to a diffusion rate 
$D_1 \sim l_1^2/t_c$ along the chain. 
For high frequency noise with amplitude  
$\epsilon$ this time is  $t_c \propto 1/\epsilon^2$\cite{Ott}.
In the case of low frequency ($\omega$) noise $t_c \propto   
(\epsilon\omega)^{2/3}$ \cite{Borg2}.
Here we only analyzed the case of high frequency noise on TIP 
problem. 

We studied the model of two interacting kicked rotators in presence of 
noise. The evolution operator is given by :

\begin{equation}
\hat{S} = \exp\{ -i[H_0(\hat{n}_1)+H_0(\hat{n}_2) +U\delta_{n_1,n_2}]\}
\times \exp\{-i[V(\theta_1,t)+V(\theta_2,t)]\}
\label{evolop}
\end{equation}

where
 $\hat {n}_{1,2} = -i\partial/\partial\theta_{1,2}$,
 $H_0(\hat n) = T {\hat n}^2 /2$, 
 $V(\theta,t) = [k + \epsilon f(t)]\cos \theta$, and
 $f(t)$ is a random function of $t$ homogeneously distributed in the 
 interval $[-1,1]$.

 For $U=0$ and $\epsilon =0$ 
 we have two decoupled kicked rotators and in the chaos domain 
 $kT > 1 $ the localization length is $l_1 \approx k^2/2$ \cite{DLS87}.
 In the presence of noise $\epsilon > 0$ the decoherence time is
 $t_c \sim 1/\epsilon^2 $. If this time is less than the localization time 
 $t^*_1 \approx l_1$   localization effects are not important and
 diffusion goes with the usual classical rate $ D \approx k^2/2$.
 On the other side when $t^*_1 < t_c$ the diffusion rate is  
 $ D_1 \sim \epsilon^2 l_1^2 $ \cite{Ott}.

 In the presence of interaction ($U\neq 0$), but without noise 
 ($\epsilon=0$), a TIP pair of size $l_1$  propagates on a 
 distance $l_c \gg l_1$ and it is localized after a time
 $t^*_2 \sim l_c l_1$ \cite{Moriond}. On a time interval 
 $t^*_1 < t < t^*_2$ the pair diffuses with a diffusion rate 
 $D_p \sim U^2 D$.
The noise leads to a destruction of localization after a 
decoherence time $t_c \sim 1/\epsilon^2$ as in the one--particle
case.
After this time the pair makes a jump of size $l_c$ 
and therefore the noise--induced pair diffusion rate can be 
estimated as $ D_+ \sim l_c^2 / t_c $ that gives 
for $t_c > t^*_2$  :

\begin{equation}
D_+ \sim \epsilon^2 l_c^2  \sim D_1 (l_c/l_1)^2  
\label{dplus}
\end{equation}

This means that for $l_c/l_1 \gg 1$ the noise--induced TIP 
diffusion rate is strongly enhanced with respect to the 
non interacting one ($D_1$).
In the case of relatively strong noise, $t_c < t^*_2$ and 
$D_+$ becomes comparable with the pair diffusion rate $D_p$
on a time scale $t^*_1  < t < t^*_2$.

The estimate (\ref{dplus}) gives the diffusion rate of the center 
of mass of the TIP pair. However 
noise also leads to a separation of two particles.
This separation goes in a diffusive way with a diffusion rate
$D_1$ which is independent from interaction $U$.
Due to this on  asymptotically large times the pair propagation
will be subdiffusive.
iAt this time the spreading of the center of 
mass $\Delta n_+^2$ can be estimated 
in the following way  
$ \Delta n_+^2 \sim \nu D_+ t$ where $\nu $ is the probability of collision 
between
two particles $ \nu \sim l_1/\Delta n_-$. Here $\Delta n_-$ is the 
effective pair size at the time $t$ which in turn can be estimated as
$ \Delta n_-^2 \sim D_1 t$. Therefore we have  
$ \Delta n_+^2 \sim (l_c/l_1)^2  l_1 \sqrt{D_1 t}$.

However on short times, when $\Delta n_- \sim l_1 $ the diffusive
spreading in $n_+$ will be given by $\Delta n_+^2 \sim D_+ t $. 
Examples of these diffusive spreading are presented in Fig.1 and Fig.2 
where 
the growth of 
$\sigma_+ = (\vert n_1\vert + \vert n_2 \vert)^2/4$
and
$\sigma_- = (\vert n_1 \vert - \vert n_2 \vert)^2$
is shown as a function of time.
These results clearly show that the diffusion rate 
$D_+= \sigma_+/t$ is 
strongly enhanced with the interaction switching on 
(approximately 20 times from $U=0$ to $U=2$).
At the same time the diffusion rate in $n_-$  ($D_1 = \sigma_-/t$)
remains practically the same (see Fig.2).
For sake of comparison we also present in  Figs. 1,2 TIP localization
in absence of noise , when $\sigma_\pm$ are oscillating in time 
near their asymptotic values.

To check the relation (\ref{dplus}) we determined the diffusion rate
$D_+$ and checked its dependence on parameters.  
According to (\ref{dplus}) 
$D_+ \sim \epsilon^2 \sigma_+^0 $ where $\sigma_+^0 \sim l_c^2 $ 
is the asymptotic 
value  of $\sigma_+$ in absence of noise.
The dependence of $D_+/\epsilon^2$ on  $\sigma_+^0$
is shown in Fig. 3 for different $\epsilon, U$ and $k$ values.
The average behaviour is given by the approximate relation
$ D_+ = 13 \epsilon^2 \sigma_+^0 $ and it is in agreement with the
theoretical prediction (\ref{dplus}).

In real systems, noise can appear as the result of electron 
interaction with phonons at finite temperature.
Our results indicate that the noise produced by phonons can lead
to a strong enhancement of diffusion (conductance) 
of electrons in a random
potential. Of course in the analysis of a physical model 
the case of finite particles 
(or quasi--particles)
density should be considered. 
In this case the probability to find two particles within a distance 
$l_1$ one from each other is of the order of 
${\cal W} \sim l_1 \rho$ where
$\rho$ is the linear (per unit length) density of particles.
This leads to a decrease of the effective diffusion rate which in 
this case can be estimated as 
$ D_{eff} \sim {\cal W} D_+ \sim D_1 l_c^2 \rho/l_1 $.
Even for small density $D_{eff}$ can be larger than $D_1$ 
and we expect that the effect of interaction--enhanced diffusion
is physically relevant. 
However future investigations on the finite particles density case
should be done.

We acknowledge the University of Como for hospitality  
during the final stage of this work.

\vfill\eject
\renewcommand{\baselinestretch} {2}

\vfill\eject

{\bf {Figure captions} }
\vskip 20pt
\begin{description} 
{
\item[Fig. 1:]
Dependence of $\sigma_+$ on time for different $U$ values:
$U=0,1,2$ correspond to lower, middle and upper full curves,
$\epsilon = 0.02, k=4, kT=5$. Dotted line is for $\epsilon=0, U=2,
k=4, kT=5$. Initially particles are at $n_1 = n_2 = 0$. The basis is
$-250 < n_{1,2} < 250$.

\item[Fig. 2:]
The same as Fig.1 but for $\sigma_-$.

\item[Fig. 3:]
Dependence of TIP pair diffusion rate on $\sigma_+^0$.
Full circles are for $k=4, \epsilon=0.01, 0\leq U \leq 2$,
open circles for $k=4, \epsilon=0.02, 0\leq U \leq 2$,
full squares for $k=5.7, \epsilon=0.01, U=0,1$, open
squares for $k=4.8, \epsilon=0.02, U=1,2$,
up-triangles for $k=4.8, \epsilon=0.01, U=1,2$
down-triangles for $k=3.3, \epsilon=0.01, U=1,2$,
diamonds for $k=3.3, \epsilon=0.03, U=1,2$.
asterisks  for $k=4, \epsilon=0.05, U=0,1$. In all cases
the chaos parameter has been fixed $kT=5$.
Dashed line shows average dependence $D_+ = 13 \epsilon^2 \sigma_+^0$.
}
\end{description}
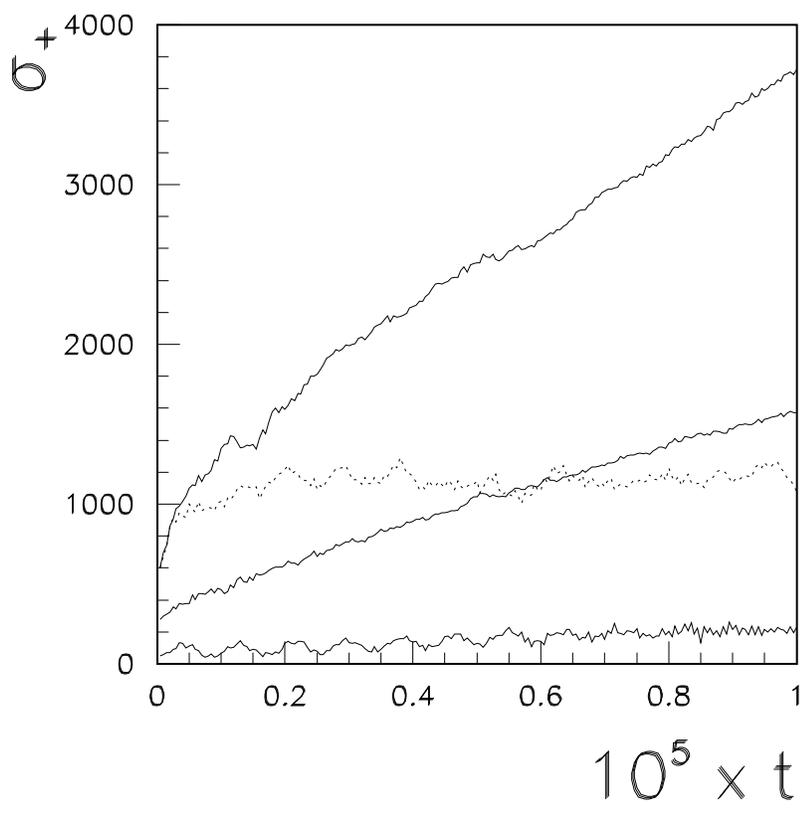
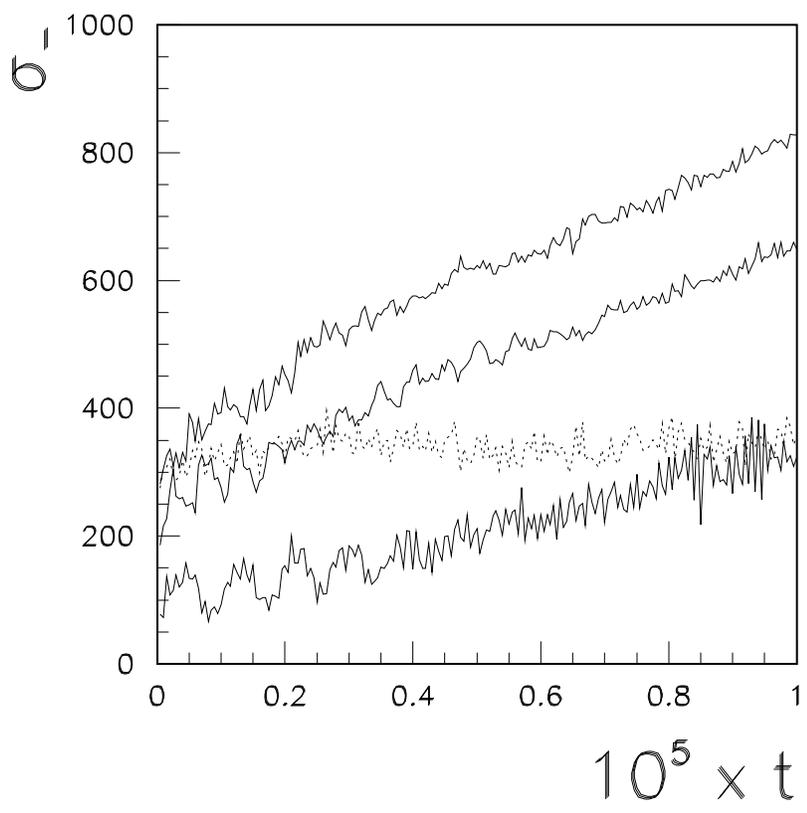
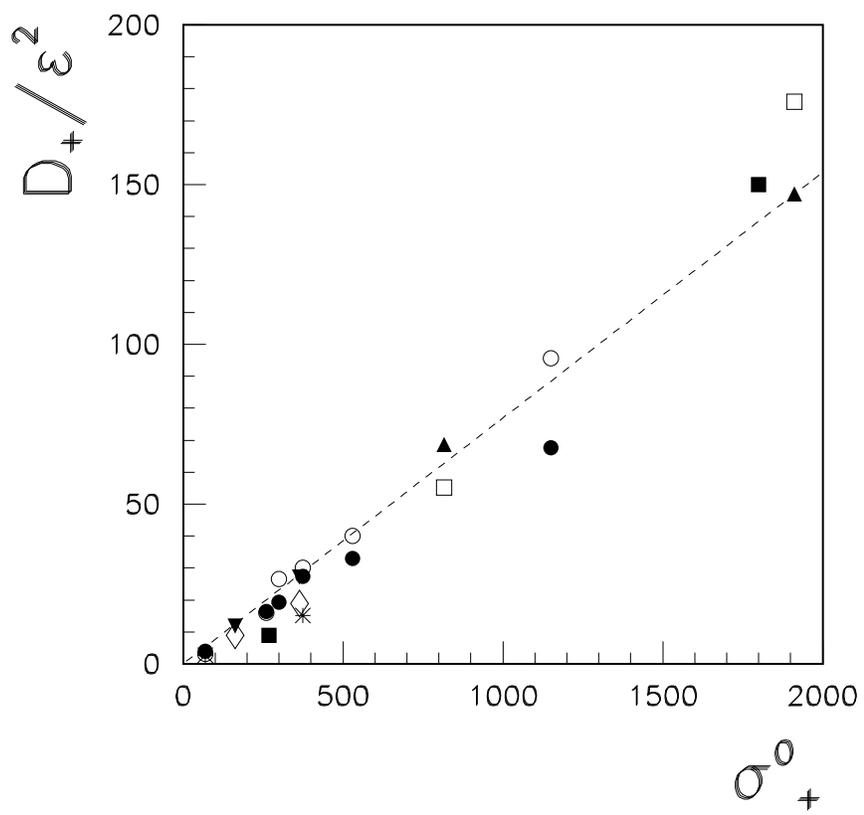
\end{document}